# Recurrence analysis of the Portevin-Le Chatelier effect

A. Sarkar[a*], Charles L. Webber Jr.[b], P. Barat[a], P. Mukherjee[a]

[a] Variable Energy Cyclotron Centre, 1/AF Bidhan Nagar, Kolkata 700064, India

[b] Department of Physiology, Loyola University Medical Center, 2160 S. First Avenue, Maywood, IL 60153, USA

**Abstract**

Tensile tests were carried out by deforming polycrystalline samples of Al-2.5%Mg alloy at room temperature in a wide range of strain rates where the Portevin-Le Chatelier (PLC) effect was observed. The experimental stress-time series data have been analyzed using the recurrence analysis technique based on the Recurrence Plot (RP) and the Recurrence Quantification Analysis (RQA) to study the change in the dynamical behavior of the PLC effect with the imposed strain rate. Our study revealed that the RQA is able to detect the unique crossover phenomenon in the PLC dynamics.

**Introduction**

The Portevin-Le Chatelier (PLC) effect is one of the widely studied metallurgical phenomena, observed in many metallic alloys of technological importance [1-12]. It is a striking example of the complexity of the spatiotemporal dynamics, arising from the collective behavior of dislocations. In uniaxial loading with constant imposed strain rate, the effect manifests itself as a series of serrations (stress drops) in the stress-time or strain curve. Each stress drop is associated with the nucleation of a band of localized plastic deformation, often designated as PLC band, which under certain conditions propagates along the sample. The microscopic origin of the PLC effect is the dynamic strain aging (DSA) [13-19] of the material due to the interaction between mobile dislocations and diffusing solute atoms. At the macroscopic scale, this dynamic strain aging leads to a negative strain rate sensitivity (SRS) of the flow stress and makes the plastic deformation nonuniform.

---

[*] Corresponding author: apu@veccal.ernet.in

In polycrystals three types of the PLC effect are traditionally distinguished on the qualitative basis of the spatial arrangement of localized deformation bands and the particular appearance of deformation curves [20, 21]. Three generic types of serrations: type A, B and C occur depending on the imposed strain rate. For sufficiently large strain rate, type A serrations are observed. In this case, the bands are continuously propagating and highly correlated. The associated stress drops are small in amplitude [22,23]. If the strain rate is lowered, type B serrations with relatively larger amplitude occur around the uniform stress strain curve. These serrations correspond to intermittent band propagation. The deformation bands are formed ahead of the previous one in a spatially correlated manner and give rise to regular surface markings [22,23]. For even smaller strain rate, bands become static. This type C band nucleates randomly in the sample leading to large saw-tooth shaped serration in the stress strain curve and random surface markings [22,23].

From metallurgical point of view, the PLC effect is usually undesirable since it has detrimental influences like the loss of ductility and the appearance of surface markings on the specimen. Beyond its importance in metallurgy, the PLC effect is an epitome for a general class of nonlinear complex systems with intermittent bursts. The succession of plastic instabilities shares both physical and statistical properties with many other systems exhibiting loading-unloading cycles e.g. earthquakes. PLC effect is regulated by interacting mechanisms that operate across multiple spatial and temporal scales. The output variable (stress) of the effect exhibits complex fluctuations which contains information about the underlying dynamics.

The PLC effect has been extensively studied over the last several decades with the goal being to achieve a better understanding of the small-scale processes and of the multiscale mechanisms that link the mesoscale DSA to the macroscale PLC effect. The technological goal is to increase the SRS to positive values in the range of temperatures and strain rates relevant for industrial processes. This would ensure material stability during processing and would eliminate the occurrence of the PLC effect.

Due to a continuous effort of numerous researchers, there is now a reasonable understanding of the mechanisms and manifestations of the PLC effect. A review of this field can be found in Ref. [17,18]. The possibility of chaos in the stress drops of PLC effect was first predicted by G. Ananthakrishna *et. al.* [24] and latter by V. Jeanclaude *et. al.* [25]. This prediction generated a new enthusiasm in this field. In last few years, many statistical and dynamical studies have been carried out on the PLC effect [10-12, 26-32]. Analysis revealed two types of dynamical regimes in the PLC effect. At medium strain rate (type B) chaotic regime has been demonstrated [30, 33], which is associated with the bell-shaped distribution of the stress drops. For high strain rate (type A) the dynamics is identified as self organized criticality (SOC) with the stress drops following a power law distribution [33]. The crossover between these two mechanisms has also been a topic of intense research for the past few years [29,33,34-36]. It is shown that the crossover from the chaotic to SOC dynamics is clearly signaled by a burst in multifractality [29,33].

This crossover phenomenon is of interest in the larger context of dynamical systems as this is a rare example of a transition between two dynamically distinct states. Chaotic systems are characterized by the self similarity of the strange attractors and sensitivity to initial conditions quantified by fractal dimension and the existence of a positive Lyapunov exponent, respectively. On the contrary, the SOC dynamics is characterized by infinite number of degrees of freedom and a power law statistics.

The general consensus that the dynamic strain aging is the cause behind the PLC effect suggests a discrete connection between the stress fluctuation and the band dynamics. We do not have a system of primitive equations to describe the dynamics of the band, so we must extract as much information as possible from the data itself. We use the stress data recorded during the plastic deformation for our analysis. However, we do not analyze these data blindly but in the framework of nonlinear dynamics as the band dynamics shows intermittency. In this work, we have carried out detailed recurrence analysis of the stress time data observed during the PLC effect to study the change in the dynamical behavior of the effect with the imposed strain rate.

**Experimental details:**

Substitutional Aluminum alloys with Mg as the primary alloying element are model systems for the PLC effect studies. These alloys have wide technological applications due to their advantageous strength to weight ratio. They show good ductility and can be rolled to large reductions and processed in thin sheets and are being extensively used in beverage packaging and other applications. However, the discontinuous deformation behavior of these alloys at room temperature rule them out from many important applications like in the automobile industry. These alloys exhibit the PLC effect for wide range of strain rates and temperatures. Under these conditions the deformation of these materials localize in narrow bands which leave undesirable band-type macroscopic surface markings on the final products.

Tensile tests were conducted on flat specimens prepared from polycrystalline Al-2.5%Mg alloy. Specimens with gauge length, width and thickness of 25, 5 and 2.3 mm, respectively were tested in an INSTRON (model 4482) machine. All the tests were carried out at room temperature (300K) and consequently there was only one control parameter, the applied strain rate. To monitor closely its influence on the dynamics of the PLC effect, strain rate was varied from $7.98 \times 10^{-5}$ $S^{-1}$ to $1.60 \times 10^{-3}$ $S^{-1}$. The PLC effect was observed through out the range. The stress-time response was recorded electronically at periodic time intervals of 0.05 seconds. Fig. 1 shows the observed PLC effect in a typical stress-strain curve for strain rate $1.20 \times 10^{-3}$ $S^{-1}$. The stress data shows an increasing trend due to the strain hardening effect. The trend is eliminated and analyses reported in this study are carried out on the resulting data. The inset in the Fig. 1 shows a typical segment of the trend corrected stress-strain curve. In the varied strain rate region we could observe type B, B+A and A serrations as reported [20,21]. We kept the sampling rate same for all the experiments. Consequently the number of data points was not same for different strain rate experiments. To analyze the data in similar footing we have carried our analysis on the stress data from the same strain region 0.02-0.10 for all strain rate experiments.

**Recurrence Analysis**

Eckman, Kamphorst and Ruelle [37] proposed a new method to study the recurrences and nonstationary behaviour occurring in dynamical system. They designated the method as "recurrence plot" (RP). The method is found to be efficient in identification of system properties that cannot be observed using other conventional linear and nonlinear approaches. Moreover, the method has been found very useful for analysis of nonstationary system with high dimension and noisy dynamics. The method can be outlined as follows: given a time series $\{x_i\}$ of N data points, first the phase space vectors $u_i=\{x_i, x_{i+\tau}, \ldots, x_{i+(d-1)\tau}\}$ are constructed using Taken's time delay method. The embedding dimension ($d$) can be estimated from the false nearest neighbor method. The time delay ($\tau$) can be estimated either from the autocorrelation function or from the mutual information method. The main step is then to calculate the N×N matrix

$$R_{i,j} = \Theta(\varepsilon_i - \|\vec{x_i} - \vec{x_j}\|), \qquad i,j=1,2,\ldots,N \qquad (14)$$

where $\varepsilon_i$ is a cutoff distance, $\|..\|$ is a norm (we have taken the Euclidean norm), and $\Theta(x)$ is the Heavyside function. The cutoff distance $\varepsilon_i$ defines a sphere centered at $\vec{x_i}$. If $\vec{x_j}$ falls within this sphere, the state will be close to $\vec{x_i}$ and thus $R_{i,j}=1$. The binary values in $R_{i,j}$ can be simply visualized by a matrix plot with color black (1) and white (0). This plot is called the recurrence plot.

However, it is often not very straight forward to conclude about the dynamics of the system from the visual inspection of the RPs. Zbilut and Webber [38,39] developed the recurrence quantification analysis (RQA) to provide the quantification of important dynamical aspects of the system revealed through the plot. The RQA proposed by Zbilut and Webber is mostly based on the diagonal structures in the RPs. They defined different measures, the recurrence rate (REC) measures the fraction of black points in the RP, the determinism (DET) is the measure of the fraction recurrent points forming the diagonal line structure, the maximal length of diagonal structures ($L_{max}$), the entropy (Shannon entropy of the line segment distributions) and the trend (measure of the paling of recurrent points away from the central diagonal). These variables are used to detect the transitions in the time series. Recently Gao [40] emphasized the importance of the vertical structures in RPs and introduced a recurrence time statistics corresponding to the vertical structures in RP. Marwan et. al. [41] extended Gao's view and defined measures of complexity based on the distribution of the vertical line length. They introduced three new RP based measures: the laminarity, the trapping time (TT) and the maximal length of the vertical structures ($V_{max}$). Laminarity is analogous to DET and gives the measure of the amount of vertical structure in the RP and represents laminar states in the system. TT contains information about the amount as well as the length of the vertical structure. Applying these measures to the logistic map data they found that in contrast to the conventional RQA measures, their measures are able to identify the laminar states i.e. chaos-chaos transitions. The vertical structure based measures were also found very successful to detect the laminar phases before the onset of life-threatening ventricular tachyarrhythmia [41]. Here we have applied the measures proposed by Marwan et. al. along with the traditional measures to find the effect of strain rate on the PLC effect.

**Results and Discussions**

RP and RQA have been successfully applied to diverse fields starting from Physiology to Econophysics in recent years. A review of various applications of RPs and RQA can be found in the recent article by Marwan et. al. [42]. Here we extend the list of application of RPs and RQA and for the first time apply these methods to study the dynamical behavior of the PLC effect in Al-2.5%Mg alloy. In this study, we particularly concentrate on the strain rate region of $7.98\times10^{-5}$ $S^{-1}$ to $1.60\times10^{-3}$ $S^{-1}$. The main goal is to demonstrate the ability of RQA to detect the unique crossover phenomenon observed in the PLC dynamics.

It has been shown that a RP analysis is optimal when the trajectory is embedded in a phase space reconstructed with an appropriate dimension $d$ [38]. Such a dimension can be well estimated using a false nearest neighbor technique. The $d$-dimensional phase space is then reconstructed using delay coordinate. The time delay $\tau$ can be estimated using the mutual information or the first zero of an auto correlation function. Based on the false nearest neighbor method we have chosen $d$ to be 10 for all the strain rates. $\tau$ obtained from the mutual information were in the range 1-14 for different strain rate data. A parameter specific to the RP is the cutoff distance $\varepsilon_i$. $\varepsilon_i$ is selected from the scaling curve of REC vs, $\varepsilon_i$ as suggested in the literature [43]. Fig. 2 shows the RPs of the stress fluctuations during the PLC effect at four different strain rates. From the visual inspection of the RPs it is easy to understand that the dynamical behavior of the PLC effect changes with the strain rate. However, it is wise to go for the RQA and quantify the difference in the PLC dynamics with strain rate. Fig. 3 shows the variation of the various RQA variables with strain rate. It can be seen from the Fig. 3 that the RQA variables like DET and laminarity do not show any systematic variation with strain rate. $L_{max}$, TT and $V_{max}$ decreased rapidly with strain rate and reached a plateau. Trend values remained almost constant at lower strain rates and decreased at higher strain rates. The variation of entropy with strain rate is rather interesting. The entropy initially decreased with strain rate and suddenly reached a higher value. However, the most important behavior was observed in the variation of REC and a variable derived from REC and DET, i.e. the ratio of DET and REC (DET/REC). The REC values decreased initially and reached a low value and then again started increasing. This variation is quite appealing in the sense that the REC value is very low in the crossover region and hence is able to detect the crossover phenomenon of the PLC effect. On the contrary, the DET/REC values showed an abrupt jump in the crossover region.

It is clearly evident from this study that RQA is able to detect the crossover in the PLC dynamics from type B to type A region. However the detailed explanation of the results obtained from the study are not straightforward. Further study is necessary which will in turn also help to understand the dislocation dynamics involved in the PLC effect.

**Conclusions**

In conclusion, for the first time we have applied the recurrence analysis to study the dynamical behavior of the PLC effect. The study revealed that the recurrence analysis is efficient to detect the unique crossover, as indicated in the earlier studies, in the dynamics of the PLC effect.

Fig. 1 True stress vs. true strain curve of Al-2.5%Mg alloy deformed at a strain rate of $1.20\times10^{-3}$ $S^{-1}$. The inset shows a typical segment of the trend corrected true stress vs true strain curve.

Fig. 2 Recurrence plots at the strain rate (a) $1.60\times10^{-3}$ $S^{-1}$ (b) $7.97\times10^{-4}$ $S^{-1}$ (c) $3.85\times10^{-4}$ $S^{-1}$ (d) $1.99\times10^{-4}$ $S^{-1}$

Fig. 3 Variation of the Recurrence Quantification Analysis variables with strain rate.

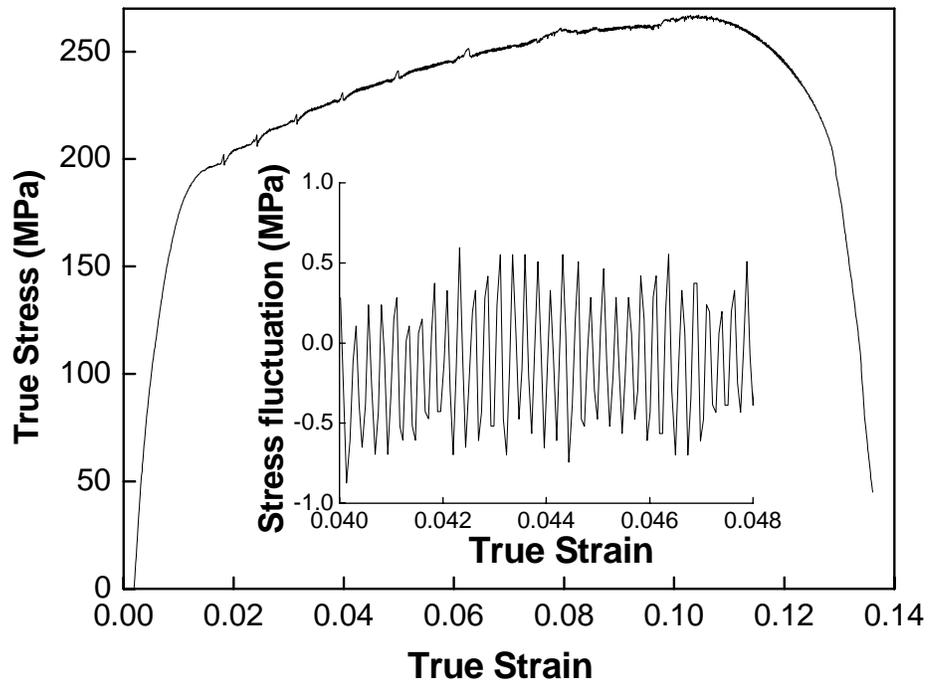

Fig. 1

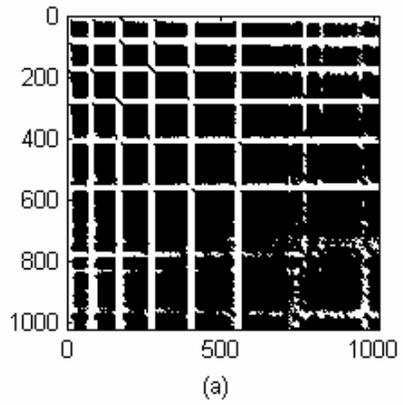 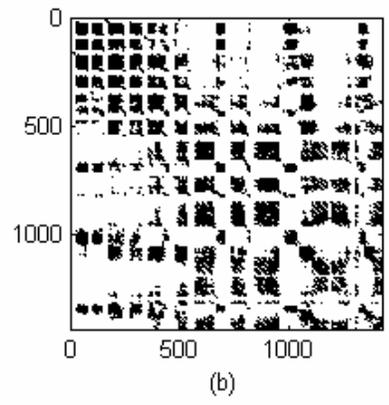
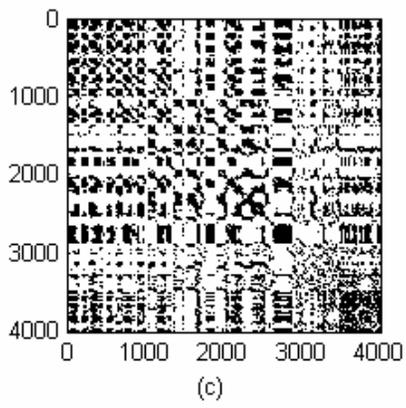 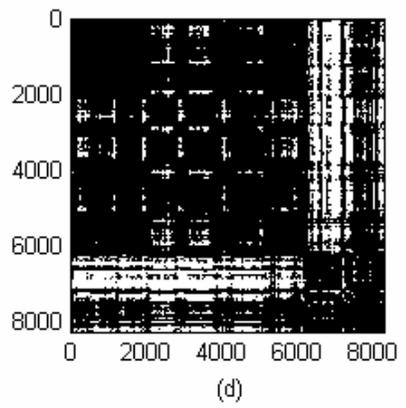

Fig. 2

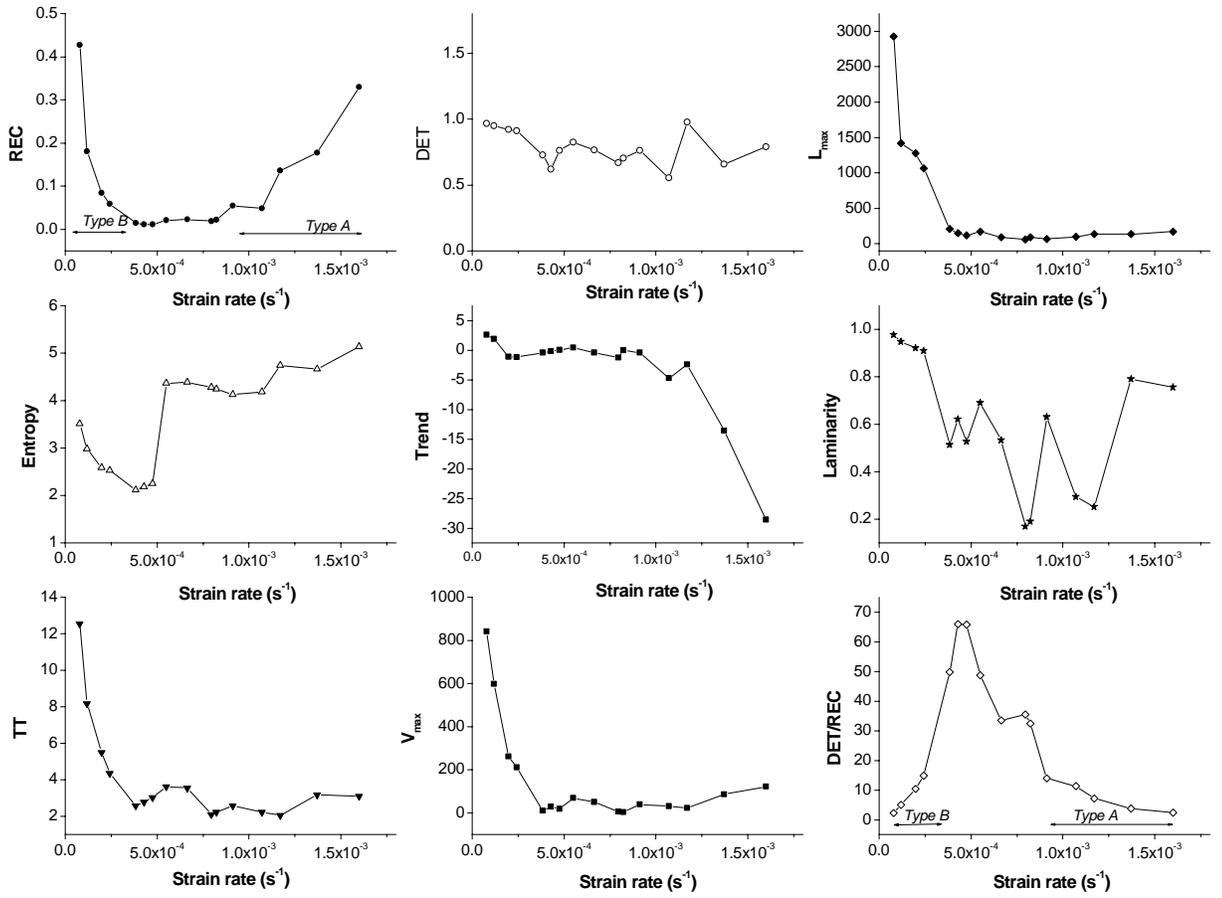

Fig. 3